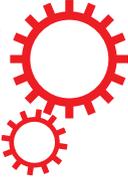

# SCIENTIFIC REPORTS

# Optical nanoscopy of transient states in condensed matter

F. Kuschewski[1], S.C. Kehr[1], B. Green[2], Ch. Bauer[2,3], M. Gensch[2] & L.M. Eng[1]

Recently, the fundamental and nanoscale understanding of complex phenomena in materials research and the life sciences, witnessed considerable progress. However, elucidating the underlying mechanisms, governed by entangled degrees of freedom such as lattice, spin, orbit, and charge for solids or conformation, electric potentials, and ligands for proteins, has remained challenging. Techniques that allow for distinguishing between different contributions to these processes are hence urgently required. In this paper we demonstrate the application of scattering-type scanning near-field optical microscopy (s-SNOM) as a novel type of nano-probe for tracking transient states of matter. We introduce a sideband-demodulation technique that allows for probing exclusively the stimuli-induced change of near-field optical properties. We exemplify this development by inspecting the decay of an electron-hole plasma generated in SiGe thin films through near-infrared laser pulses. Our approach can universally be applied to optically track ultrafast/-slow processes over the whole spectral range from UV to THz frequencies.

Understanding the complexity of solid-state materials such as high-temperature superconductors, functional proteins and two-dimensional materials has become a subject of intense research in the last 20 years. Nevertheless, these systems still present challenges for standard spectroscopic or imaging analysis since they exhibit an entanglement of different degrees of freedom on the mesoscopic to atomistic length scale[1–3]. In order to further increase the knowledge on these systems and make profit for valuable applications, transient-state probing by applying different means and technologies is one of the key problems to solve[4]. It has been shown that some of the underlying mechanisms and phenomena occurring in such complex systems in fact may be disentangled through their characteristic response times, for instance when applying specially selected external stimuli[5,6]. A second important diagnostic advancement was established through developing versatile microscopy and imaging techniques that allow for local probing at different length scales, down to the single domain/single molecule or even sub-Angstrom level[7–13].

Techniques that provide both a high spatial and temporal resolution while simultaneously being sensitive to different transient states are rare. One method that has recently been reported is time-resolved X-ray scattering and diffraction (XAS) using ultra-short X-ray pulses as delivered e.g. by X-ray free-electron lasers[14,15]. Due to the rather limited accessibility to such large-scale facilities only a few reports exist to date, however, clearly demonstrating the benefits and advantages when probing temporal and spatial properties close to the quantum limit[16]. Such techniques hence are in the focus of various developments ranging from ultra-fast time-resolved electron or improved X-ray diffraction/scattering[17], to versatile time-resolved electron microscopy[18–21]. Also, recent studies demonstrate both an ultra-high spatial and temporal resolution by combining scattering-type scanning near-field optical microscopy (s-SNOM) with pulsed THz-light sources and time-resolved all-optical microscopy[22,23]. In all these works the availability of sophisticated laser light sources providing high duty cycles was mandatory for the successful application of these methods.

The approach that we follow in this paper allows for probing the optical response of a transient state of matter on the nanometer length-scale providing a pico- to femtosecond time resolution via recording the local changes of the near-field optical properties. We use a custom-made s-SNOM[11,24], in combination with a novel sideband demodulation technique for detection. The novel method is demonstrated on a photo-induced electron-hole plasma generated at the surface of a SiGe thin film, and demonstrates for the first time that changes in the optical constants between two transient states can be probed optically with a sub-diffraction-limited spatial resolution. Moreover, we prove that our technique is operable with a manifold of pulsed laser light sources, such as a table-top Nd:YAG laser and also an infrared free-electron laser (FEL). In the latter case, our technique allows for accessing the local optical response over a broad frequency range (tunable from 1.2 to 80 THz) and with Fourier-limited temporal resolution in the ps regime[25]. Similar to macroscopic pump-probe experiments this resolution is determined by the pulse duration of stimulus and probe and by their temporal synchronization. Hence, combining our technique with ultra-fast laser-driven sources[26–28], thus must result in even an improved time resolution down to 100 fs and below. Most importantly our method requires neither high duty cycles nor high repetition rates of the stimuli, nor is it confined to transient states driven by photoexcitation. It therefore may be easily employed to study a huge range of fundamental physical processes across very different time scales, i.e. from microseconds to several 10 seconds. The latter regime is of great importance, for instance, in the life sciences[6,29].

[1]Institute of Applied Physics, Technische Universität Dresden, Dresden, Germany. [2]Institute of Radiation Physics, Helmholtz-Zentrum Dresden-Rossendorf, Dresden, Germany. [3]Department of Physics, Freie Universität Berlin, Berlin, Germany. Correspondence and requests for materials should be addressed to S.C.K. (email: susanne.kehr@iapp.de)

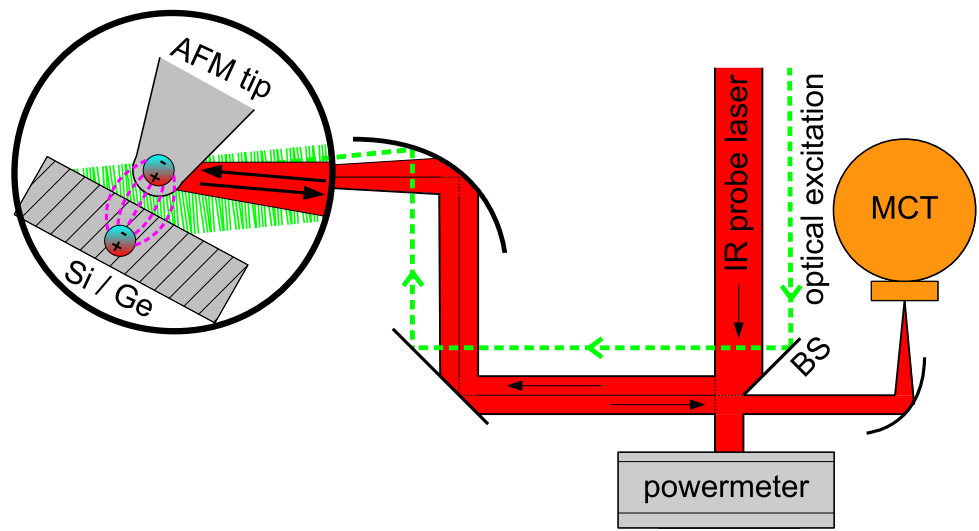

**Figure 1. Experimental setup.** The IR probe beam ($CO_2$/FEL, red) is divided into two branches using a geometrical beam splitter (BS) (Au-evaporated Si wafer), which redirects part of the beam towards the s-SNOM and, in addition, allows for *in-situ* monitoring of the laser power in transmission. The beam is focused onto the AFM tip/sample junction with the help of a parabolic mirror. The backscattered light is directed onto a MCT detector passing a germanium filter (not shown) and then demodulated via the side-band technique (see text and Fig. 2). The optical excitation using a Nd:YAG pump laser (green) is fed-in parallel to the probe beam and focused on the tip-sample system through the same parabolic mirror.

## Results

### SNOM setup and signal demodulation.
Our s-SNOM setup is based on a non-contact atomic force microscope (AFM) as described in detail in the methods section. The typical lateral optical resolution of our setup measures ~10 nm. In order to probe the transient state behavior, we employ an optical pump-probe scheme (see Fig. 1) as described in the following.

The probe lasers consist of either a tunable continuous-wave (cw) $CO_2$ laser or a free-electron laser (quasi-cw, repetition rate $f_{FEL} = 13$ MHz, pulse length $\tau_{FEL} = 1-25$ ps). Both of them are tuned to a $\lambda = 10.6 \mu m$ wavelength. The probe beam is directed towards the s-SNOM via a geometrical beam splitter and focused onto the s-SNOM probe using a parabolic mirror, leading to a probe-focus diameter of approximately 100 $\mu m$. The same parabolic mirror is used to collect the backscattered light that is measured with a standard MCT detector.

For the experiments discussed in this paper, we complemented our setup with a Nd:YAG pump laser at $\lambda = 1064$ nm with a repetition rate of $f_p = 1$ kHz and a pulse width of $\tau_p = 20$ ps. The Nd:YAG beam is coupled into the beam path in parallel to the probe beam and focused to the pump-sample system with a focus size of 1 mm. In order to avoid stray light from the pump beam to influence the measurement a germanium filter is mounted in front of the detector.

Generally, s-SNOM constitutes a non-invasive and non-destructive scanning probe method that allows for the examination of local optical constants with a wavelength-independent lateral resolution on the order of 10 nm. For scattering-type near-field optics, the nanoscopic s-SNOM probe transforms the local near-field information of the sample into a propagating wave that reaches the detector in the far field. In order to differentiate the subtle near-field information from the typically much larger far-field scattering, we apply the method of higher-harmonic demodulation that utilizes the strongly non-linear dependence of the evanescent near field[30–32], (see Fig. 2a,b). For cantilever oscillation amplitudes much smaller than the wavelength, demodulation at $2\Omega$ (with $\Omega$ the fundamental cantilever oscillation frequency) is typically sufficient to extract the pure near-field signals[33]. Moreover, the near-field response can be enhanced[11,34,35] whenever the sample is excited resonantly, which is typically the case for sample permittivity values $\mathrm{Re}(\varepsilon) = -5$ to $-1$. Here, the near-field coupling of tip and sample is maximized while the signal-to-noise ratio (SNR) is dramatically increased, hence being very sensitive to small changes of the optical sample properties.

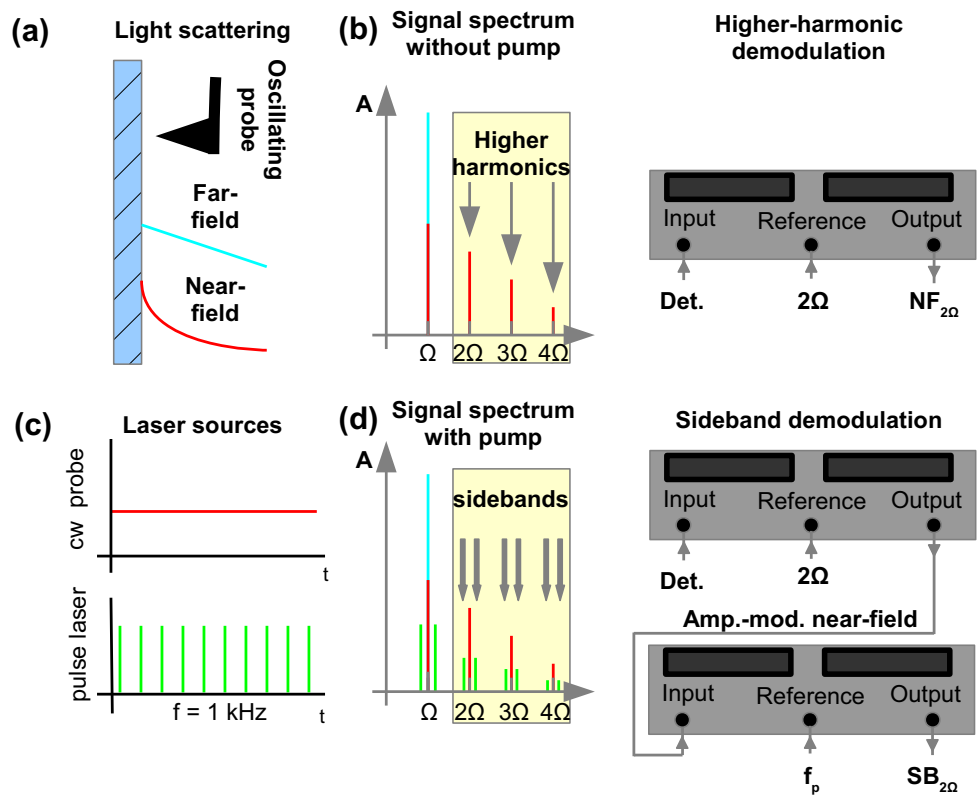

**Figure 2. Signal demodulation.** (**a**) The scattered signal contains both a large far field and a small near-field signal. The probe oscillates above the sample at the fundamental oscillation frequency $f_t \approx 150$ kHz with a 50 nm oscillation amplitude. (**b**) In contrast to the quasi-linear behavior of the far-field signal, the evanescent near field decays strongly non-linear with the sample distance. Hence, when demodulating the scattered light in individual Fourier components, the first harmonic contains both far-field and near-field signals, while higher harmonics reflect pure near-field contributions. In the ground state (no sample pumping) higher-harmonic demodulation at $2\Omega$ is realized using one LIA only in order to separate the near field from the far-field signal. (**c**) Temporal behavior of the laser sources. (**d**) When pumping the sample, sidebands are induced on all harmonics. In order to quantify their strength, a first LIA is used at $2\Omega$ yielding an amplitude-modulated near-field signal. Subsequently, a second LIA acquires the sideband information carrying the pump-induced modification of the near-field signal.

We analyze here the near-field signal with a two-fold demodulation approach. On the one hand, we extract the near fields by filtering the scattered signal with a lock-in amplifier (LIA) at the second harmonic ($2\Omega$) frequency of the fundamental cantilever oscillation $\Omega$[31,32] (see Fig. 2a,b). The $2\Omega$ near-field signal ($NF_{2\Omega}$) reflects the time-averaged near-field response of the sample carrying information on the local optical properties of the sample region probed by the AFM tip apex.

In a second step, we enhance our method by a novel two lock-in demodulation technique that measures the photo-induced sideband of the transient near-field signal at the corresponding harmonic (see Fig. 2c,d and also supplementary Fig. S1). Here, the first LIA is centered on the $2\Omega$ frequency as well, in order to separate near- from far-field contributions. However, upon sample excitation with the pulsed laser, the output signal shows an amplitude-modulated near-field behavior exhibiting sidebands left and right of the $2\Omega$ carrier frequency at multiples of the photoexcitation frequency, as schematically depicted in Fig. 2d and shown by numerical simulations in Fig. S1. Hence, we use a second LIA to demodulate this sideband information ($SB_{2\Omega}$), which then contains the desired transient photo-induced near-field component of our sample. Details of the lock-in amplifier settings used for this technique are given in the methods section.

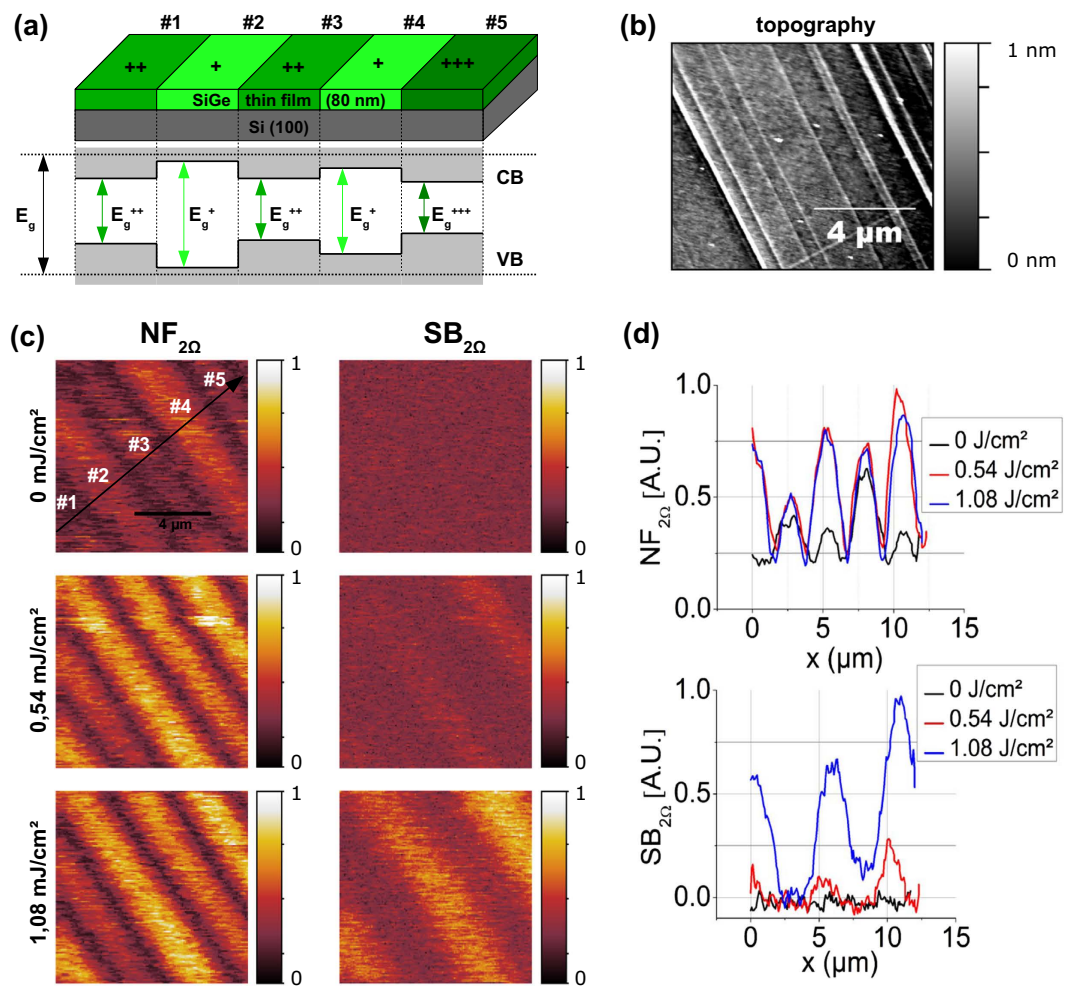

**Figure 3. Scans of Si-Ge thin-film.** (**a**) Sample sketch of a 80 nm Si(23%)-Ge(77%) film on Silicon (100). Due to lattice mismatch, the mixed-phase top layer experiences a strong stress during epitaxial growth. Dislocations and impurities partially relieve the stress in the thin film leading to the stripe-like cross-hatched structure as seen in the sample topography[38] (see (**b**)) with varying stress on different stripes. Moreover, as the stress induces band shifting, the bandgap on different sample stripes varies, affecting the absorption cross-section when pumping the SiGe sample with a pulsed laser (as sketched in (**a**)). (**c**) Near-field $NF_{2\Omega}$ (left) and sideband $SB_{2\Omega}$ (right) scans of the sample for different pump fluences. (**d**) Cross-sections taken along the line in (**c**) illustrating the stimuli-induced changes in signal intensity of $NF_{2\Omega}$ (top) and $SB_{2\Omega}$ (bottom) for different fluences.

**Electron-hole-plasma excitation in SiGe thin films.** In order to demonstrate the sideband-demodulated near-field technique as described above, we study the photo-induced near-field signature of transiently excited SiGe thin films (for details see methods section). The fundamental idea of this experiment is to probe the pump-induced electron plasma, that lives in the range of $\mu s$, with one of our two probe beams, namely either a cw $CO_2$ laser or the pulsed FEL.

The SiGe samples show a cross-hatched micro-structure[36], (Fig. 3a,b) with the different stripes having the same SiGe concentration, but experiencing a non-uniform mechanical stress that leads to unequal local optical properties. Moreover, the stress-induced optical response[37], varies appreciably due to band shifting of up to 0.1 eV/GPa for pure silicon and 0.15 eV/GPa for pure germanium. Hence, in particular when pumping with wavelengths matching the band gap of 1.07 eV, the stripes' absorptions differ dramatically. Furthermore, band warping[38–42], will also locally influence the effective mass of the charge carriers[43]. As a consequence, the stripes exhibit a non-equal photoexcitation efficiency of the electron-hole plasma leading to different charge carrier densities. When applying a Drude-type model, the impact of the induced charge carriers on the permittivity is also changing for different stripes due to the varying effective masses.

s-SNOM is sensitive to these local optical properties of the SiGe sample, and hence to any stress-induced variation of the optical constants in the sample[44]. A stress-induced contrast between the Si-Ge stripes of up to 35% can be measured already for the ground state, i.e. without any photoexcitation (see near-field signal $NF_{2\Omega}$ in Fig. 3c). Here, the optical constants of the stripes reveal values of $Re(\varepsilon) = +12$ leading to a weak non-resonant near-field signal (as compared to the resonant excitation described below).

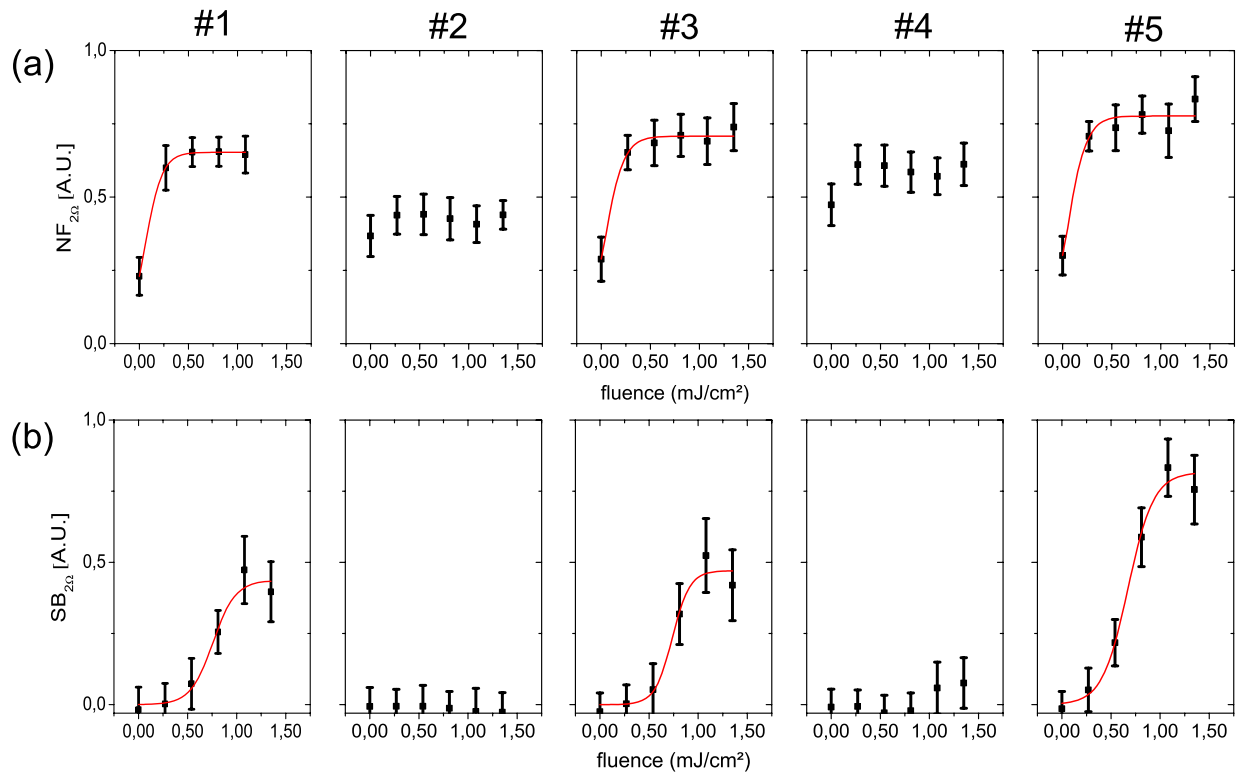

**Figure 4. Fluence dependence of the signals obtained on different stripes of the SiGe thin film.** (**a**) Integral near-field signals as a function of pump fluence averaged along the stripes as labeled in Fig. 3(a). Without excitation, stripes #2 and #4 show the strongest near-field signal. When pumping the sample with a fluence of $0.28\,mJ/cm^2$, the obtained signal increases on stripes #1, #3 and #5 leading to a contrast reversal between the stripes. When increasing the fluence further, the effect saturates. (**b**) Without excitation the sideband signal is zero. With increasing fluence stripes #1, #3 and #5 experience a strong pump effect, matching the near-field signal. This effect saturates at a fluence of around $1\,mJ/cm^2$.

Upon photoexcitation, the optical properties of the sample change for a short time period. The duration of this effect is given by the decay time of the electron-hole plasma and was determined to be $20–40\,\mu s$ by a macroscopic optical switch experiment, similar to those described in[45–47]. During this period, the permittivity of the sample becomes negative and the sample reflectivity is increased. The relevant temporal regime of near-field resonant excitation, i.e. when $Re(\varepsilon) = -5$ to $-1$, is reached shortly after the sample pumping, and causes an intense peak in the near-field response. Hence, the time-averaged near-field signal $NF_{2\Omega}$ must increase for all sample stripes. Moreover, since the photoexcitation of the different stripes is unequal, the recorded near-field amplitudes increase by a factor of 1.1 for stripes #2 and #4, to 2.8 for stripes #1, #3, and #4, which results in a change of the near-field contrast of up to 28% (see Fig. 3c,d). This photo-induced enhancement of the near-field signal saturates at around $0.54\,mJ/cm^2$ which indicates that the permittivities of the photo-excited sample areas are already more negative under the pump fluence than needed for resonant near-field excitation.

On the contrary, the sideband signal is sensitive to the strength of the near-field modulation rather than its time-average, and consequently reflects the photo-induced contribution to the near-field signal (Fig. 3c,d). Here, no contrast is observed without photoexcitation. With increasing fluence of the Nd:YAG laser pulses, the sample properties and the resulting near-field response are modulated much stronger, and hence, the stripes clearly become discernable. Unlike the time-averaged $NF_{2\Omega}$ near-field signal, this effect does not saturate at $0.54\,mJ/cm^2$, but increases further up to at least a fluence of $1.08\,mJ/cm^2$. A SNR of up to 94% thus could be measured exceeding that of the integral near-field signal by a factor of 3.

In Fig. 4, we compare the integral near field and the sideband response with the photoexcitation, varying both the fluence and the stripe position as labeled in Fig. 3a. Without any excitation, stripes #2 and #4 show a stronger near-field signal than stripes #1, #3 and #5. When pumping the sample, stripes #1, #3 and #5 are excited much stronger than stripes #2 and #4. Hence, in the photo-excited state, a contrast reversal between the stripes is observable. The corresponding sideband measurement shows a matching behavior. Stripes #1, #3 and #5 exhibit a strong photo-induced increase of the sideband signal, whereas stripe #2 and #4 show no major effects. Moreover, stripe #5 exhibits approximately twice as much near-field response as compared to stripes #1 and #3, corresponding to an increased stress of that stripe. The sideband behavior matches well with the recorded near-field data. In fact, as the bandgap energy of SiGe is close to the photon energy of the pump laser[44], the stress effect on the band structure causes different absorption of the pump-laser for different stripes. Thus the induced charge carrier density must vary, leading to different near fields and sideband amplitudes as reported in Fig. 4.

The lateral width of the stripes' edges in both signal channels is found to measure ~1 μm and matches the reported dimension of the stress-induced superstructure in the SiGe film[48]. This length may be expressed in terms of wavelengths and reaches a value of approximately $\lambda/10$, well below the diffraction limit. Again note that the spatial resolution of our near-field microscope generally lies on the order of 10 nm.

The lifetime of the electron-hole plasma in the SiGe films is a few μs. Compared to the repetition rate of the pump laser of fp = 1 kHz, this corresponds to a duty cycle of ~$10^{-3}$. However, as discussed above the s-SNOM method is most sensitive when operated close to the resonant excitation of the sample (i.e. for Re($\varepsilon$) = −5 to −1), which is not met over the full pump cycle but for an ultra-short time period only.

In terms of responsivity to changes in the sample properties, the sideband technique yields a solely pump-related contrast and therefore clearly exceeds the abilities of basic near-field microscopy. The main advantages of the method as presented here are the sensitivity to ultra-short transient changes rather than its time-averaged contribution, as well as the non-existence of any ground-state background signal in the sideband channel; hence, the signal-to-noise ratio reaches extremely high values even for room-temperature applications as the one detailed in this article.

## Discussion

A new approach for probing transient states of matter on the nanometer length scale has been demonstrated in this work. The presented novel sideband demodulation technique allows one to separate, for the first time, transient modulated signals from the integral near-field signal. Proof-of-principle experiments were performed on SiGe thin-film samples upon optical excitation, establishing a resolution beyond the diffraction limit. The required fluence for the generation of the electron-hole plasma in semiconductors of about 1 mJ/cm² is in good agreement to macroscopic experiments quoting 1–15 mJ/cm² [45–47]. A clear spatial contrast has been observed originating from a stress-modulated local modification of the band structure[38,44].

The demonstrated concept enables scanning optical near-field microscopy to be applied to transient nanoscopy with an unprecedented sensitivity. In contrast to recently performed pilot experiments of time-resolved scanning near-field infrared microscopy[22,23,49,50], our approach allows us to detect minute changes of the near-field signal by employing sideband demodulation. Our method does not require high repetition rates or large duty cycles, and thus can be employed to study samples under the influence of a variety of different external periodic stimuli, induced e.g. by photon pulses, pressure, temperature, stress, electrical and/or magnetic fields. Moreover, we incorporated into our method the ability to operate with an infrared free-electron laser (see supplementary information, Fig. S2), proving that operation over a large spectral range down to 1.2 THz is feasible.

## Methods

**s-SNOM setup and signal demodulation.** Our s-SNOM is based on a home-built non-contact AFM with optical accessibility to the AFM tip. The tip-sample junction is illuminated under an angle of 65° with respect to the sample normal. We use commercially available platinum-iridium-coated silicon tips from Nanosensors as the scattering probes in s-SNOM, operated in non-contact mode at a frequency and amplitude of $f_t \approx 150$ kHz and A = 50 nm, respectively. The tip apex with a ~10 nm tip radius interacts locally with the sample, allowing for a resolution beyond the optical diffraction limit. The tip scatters the local near-field signal into propagating waves that are measured in the far field. Here, we detect the backscattered light with a liquid-nitrogen-cooled MCT detector (Judson Technologies J15D12). A silicon wafer having an evaporated gold film of 100 nm is used as a geometrical beam splitter in our optical setup allowing us to monitor the incident probe-laser power $P_{probe}$ while recording the scattered near fields. All signals are normalized to $P_{probe} \approx 25$ mW and were demodulated at the second harmonic of the cantilever oscillation $2\Omega$ for quantifying and differentiating the near field from any far-field signals. We define the optical contrast/visibility to V = (A−B)/(A+B) with V ranging between −1 to +1 (note that the absolute value and sign hence denote the signal strength and direction, respectively). The AFM was operated with a RHK SPM 100 + PLL Pro 2 AFM controller for data acquisition.

Parameters used for the sideband-demodulation technique are a bandwidth of 18 dB and an integration time of 300 μs of the first LIA (type: Stanford Research Systems SR844, triggered at $2\Omega$), offering enough band pass for the higher harmonic and sidebands. The second LIA (type: Stanford Research Systems SR830) triggered at the photoexcitation frequency of 1 kHz is used with an integration time of 300 ms and a filter of 24 dB. Here, the filter needs to be sufficiently sharp in order to separate the sideband from the $2\Omega$ peak, while the integration time was adjusted such to allow for fast acquisition at a reasonable signal-to-noise ratio.

**Sample preparation and optical properties.** The SiGe thin films consist of an 80 nm thick, homogeneous distribution of a SiGe mixed phase containing 77% silicon and 23% germanium which was grown epitaxially with MBE on a Si (100) substrate. In the unstrained state, optical properties are determined by the composition of the two elements. Here, typical values for optical parameters are a permittivity at $\lambda = 10.6\,\mu\text{m}$ of $\varepsilon_1 = 12.74$[51], and a bandgap of $E_g = 1.07\,\text{eV}$[48]. The charge carrier mobility in SiGe heavily depends on the doping levels as well as on the strength and direction of any induced sample strain, and typically reaches values of $\mu_e = 500–1000\,\text{cm}^2/\text{Vs}$[52], and $\mu_h \approx 800\,\text{cm}^2/\text{Vs}$ for electrons and holes[53], respectively. Due to lattice mismatch between film and substrate, thin-film SiGe samples typically exhibit a cross-hatched structure (see Fig. 3a) mostly arising from strain release.

## Acknowledgements

We acknowledge financial support by the German Research Foundation (DFG) through the Cluster of Excellence "Center of Advancing Electronics Dresden" as well as by the Open Access Publication Funds of the TU Dresden, Germany. C.B. and M.G. acknowledge support through the BMBF project no. 05K10KEB. F.K., S.C.K. and L.M.E. acknowledge support through the BMBF project no. 05K10ODB. Furthermore, F.K. acknowledges the support by the Rosa Luxemburg Foundation. Special Credit goes to W. Seidel, the ELBE team, M. Fehrenbacher and T. Kämpfe for their support and to P. Hermann for the preparation of the sample.

## Author Contributions

The project idea was developed by F.K., S.C.K., M.G. and L.M.E. The experiment was performed by F.K., S.C.K., B.G. and C.B. The manuscript text was written by F.K., S.C.K., M.G. and L.M.E. F.K. prepared all figures, including the numerical simulations. B.G. reviewed the manuscript before submission.

## Additional Information

**Supplementary information** accompanies this paper at http://www.nature.com/srep

**Competing financial interests:** The authors declare no competing financial interests.

**How to cite this article**: Kuschewski, F. *et al.* Optical nanoscopy of transient states in condensed matter. *Sci. Rep.* **5**, 12582; doi: 10.1038/srep12582 (2015).